\documentclass[fleqn,10pt]{wlscirep}
\usepackage[T1]{fontenc} 
\usepackage[english]{babel}
\usepackage{microtype} 						
\usepackage{amsmath} 							
\usepackage{amssymb}							
\usepackage{lipsum}
\usepackage{soul}

\renewcommand{\b}[1]{\mathbf{ #1}}				
\newcommand{\h}[1]{\hat{ #1}}					
\renewcommand{\d}{\mathrm{d}}			       
\newcommand{\der}[1]{\frac{\mathrm{d}}{\mathrm{d} #1}}			       
\newcommand{\ket}[1]{| #1 \rangle}				
\newcommand{\braket}[2]{\langle  #1|  #2\rangle}				
\newcommand{\m}[1]{\langle  #1 \rangle}				
\newcommand{\up}{\uparrow}				
\newcommand{\down}{\downarrow}				
\newcommand{\Tr}{\mathrm{Tr}}			       

\title{Quantum measurement-induced antiferromagnetic order and density modulations in ultracold Fermi gases in optical lattices}

\author[1,*]{Gabriel Mazzucchi}
\author[1]{Santiago F. Caballero-Benitez}
\author[1]{Igor B. Mekhov}
\affil[1]{Department of Physics, Clarendon Laboratory, University of Oxford, Parks Road, Oxford OX1 3PU, United Kingdom}

\affil[*]{gabriel.mazzucchi@physics.ox.ac.uk}

\begin{abstract}
Ultracold atomic systems offer a unique tool for understanding behavior of matter in the quantum degenerate regime, promising studies of a vast range of phenomena covering many disciplines from condensed matter to quantum information and particle physics. Coupling these systems to quantized light fields opens further possibilities of observing delicate effects typical of quantum optics in the context of strongly correlated systems. Measurement backaction is one of the most fundamental manifestations of quantum mechanics and it is at the core of many famous quantum optics experiments. Here we show that quantum backaction of weak measurement can be used for tailoring long-range correlations of ultracold fermions, realizing quantum states with spatial modulations of the density and magnetization, thus overcoming usual requirement for a strong interatomic interactions. We propose detection schemes for implementing antiferromagnetic states and density waves. We demonstrate that such long-range correlations cannot be realized with local addressing, and they are a consequence of the competition between global but spatially structured backaction of weak quantum measurement and unitary dynamics of fermions.
\end{abstract}
\begin{document}

\flushbottom
\maketitle

\thispagestyle{empty}

\section*{Introduction}

The study of quantum gases trapped in optical lattice potentials is a truly multidisciplinary field~\cite{Lewenstein}. The experimental realization of toy Hamiltonians like the Hubbard model opened the opportunity of studying intriguing many-body effects in Fermi systems such as high temperature superconductivity and quantum magnetism. The latter one is particularity challenging to observe in ultracold gases  because of the extreme cooling it requires in order to create quantum states with very low entropy which exhibit antiferromagnetic (AFM) correlations. Recent experiments succeeded in realizing these states and investigated the effect of lattice geometry and dimensionality on the magnetic correlations of the ground state of the Hubbard model~\cite{Greif2015,Hart2014}. In these setups, the presence of AFM ordering is revealed by averaging the results of time-of-flight images over many different experimental runs. Moreover, classical light beams are used for manipulating, controlling and cooling the atoms. In this work, we show that the backaction arising from global spatially structured quantum measurement allows to engineer and detect quantum states presenting AFM correlations in a single experimental realization and in real-time, even in absence of interactions between atoms with opposite spin. We achieve this by coupling the atoms to a quantized light field, focusing on the ultimate regime where the quantum properties of both light and matter are equally important. Specifically, we consider the case where atoms in an optical lattice scatter light in an optical cavity. This setup has been recently realized~\cite{Hemmerich2015,Esslinger2015}, leading to the observation of new quantum phases arising from the light-mediated interaction. We focus on the quantum properties of the scattered photons leaving the optical cavity (\cite{Mekhov2012,ritsch2013} for a review) and show how performing a global quantum measurement on the atomic system can be used for tailoring even local properties  such as density modulations and AFM correlations, without the need of local addressing~\cite{Diehl2008,DiehlAFM,Kaczmarczyk2016}. Because of the entanglement between the light and matter, measurement backaction strongly affects the evolution of the atoms competing with the typical dynamics given by the tunneling processes. In contrast to quantum nondemolition approaches~\cite{Meystre2009,Eckert2008,Mekhov2007,Javanainen2003,Aguilar2014,Rogers2014,Kozlowski2015,Elliott2015}, where either the measurement backaction or the many-body dynamics were neglected, or recent proposals~\cite{Pedersen2014}, where measurements are performed at optimized moments in time, we describe the full conditional evolution of the atomic system subjected to continuous monitoring. We show that this process establishes long-range correlations allowing to directly observe the formation of density modulations and AFM order.  Importantly, these effects are visible even in a single experimental realization and do not rely on the effective cavity potential which can lead to self-organization~\cite{ritsch2013,Kramer2014,Morigi2015,Piazza2015, Caballero2015,Caballero2015b}, including that of fermions \cite{Keeling2014superradiance,Piazza2014superradiance, Chen2014superradiance}. 

\section*{Results}
\begin{figure}[b!]
\centering
\includegraphics[width=0.6\textwidth]{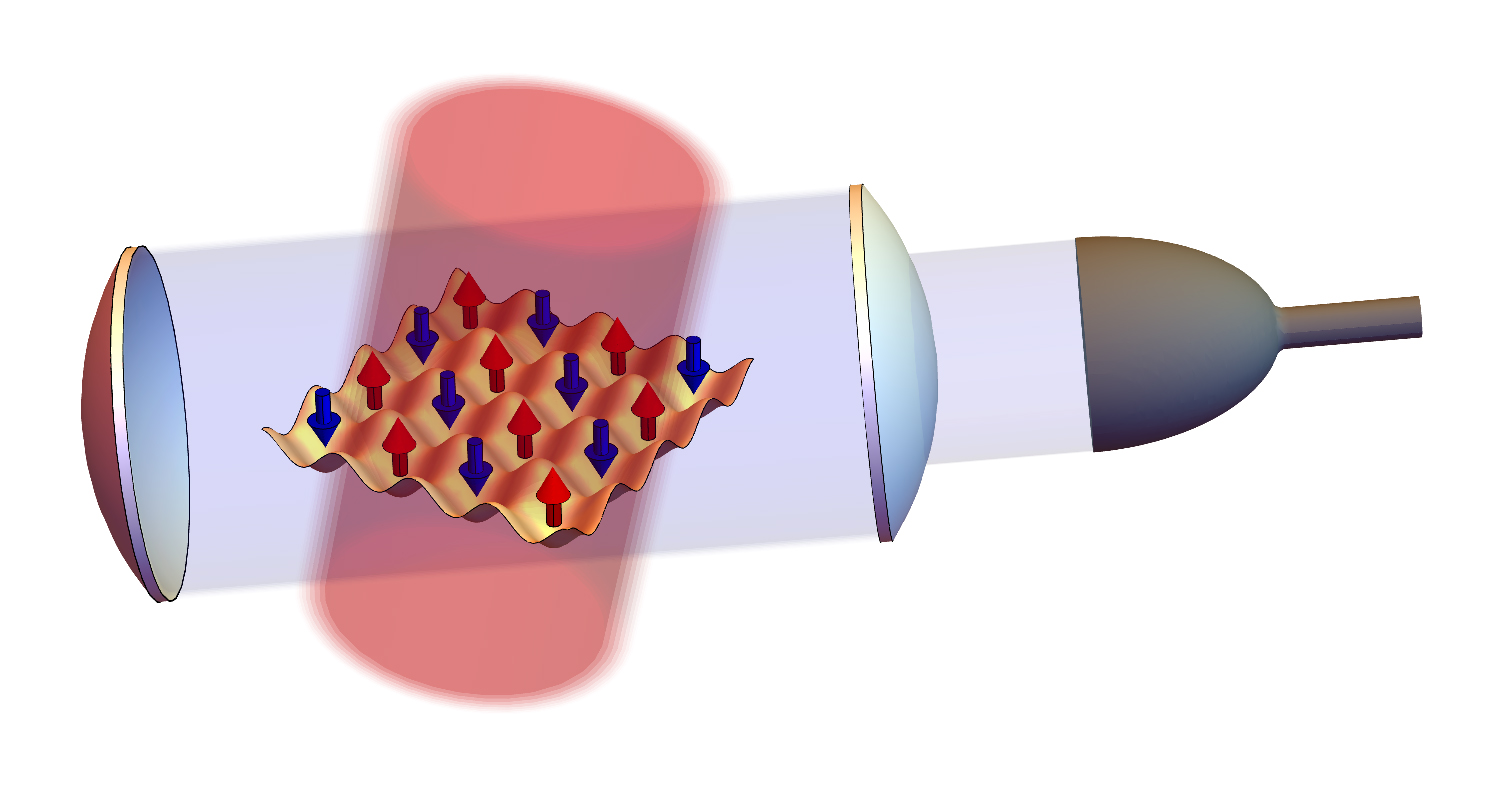}
\caption{\textbf{Experimental setup.} Ultracold fermions  are loaded in an optical lattice and probed with a coherent light beam (red). The scattered light (blue) is collected and enhanced by a leaky cavity, allowing the detection of the escaped photons.} \label{diagram}
\end{figure}

\subsection*{Theoretical model}
We consider light scattering from ultracold fermions loaded in an optical lattice with $L$ lattice sites and lattice spacing $d$. An optical cavity with decay rate $\kappa$ selects and enhances the light scattered at a particular angle~\cite{Bux2013,Kessler2014,Landig2015}. The atomic dynamics is described by the usual Hubbard Hamiltonian
\begin{equation}\label{HHubbard}
\h{H}_0 =-\hbar J \sum_{\sigma=\up,\down} \sum_{\langle i,j\rangle} \h{f}_{j,\sigma}^\dagger \h{f}_{i,\sigma} +\hbar U \sum_i \hat{n}_{i,\uparrow}\hat{n}_{i,\downarrow}
\end{equation}
where $J$ is the tunneling amplitude, $U$ the interaction energy between atoms with opposite spin and $\h{f}^\dagger_{j,\sigma}$ ($\h{f}_{j,\sigma}$) creates (annihilates) an atom with spin $\sigma$ at the lattice site $j$. The coupling between light and matter is obtained by generalizing the model in~\cite{Mekhov2012}. We introduce the polarization of the light field as an additional degree of freedom so that the light-matter Hamiltonian is
\begin{equation}\label{eq:coupling}
\h{H}_\mathrm{LA} =\hbar \sum_\sigma U_{\sigma} \h{a}_{1}^\dagger a_{0\sigma} \h{F}_{\sigma}  +\mathrm{H.c.}
\end{equation}
where $\h{a}_1$ is the annihilation operator for the cavity mode, $a_{0}$ is a classical coherent probe (Figure~\ref{diagram}), $\sigma$ is an index describing the light polarization in the circular polarization basis ($L$ or $R$), $U_{\sigma}=g_{1\sigma}^*g_{0\sigma}/\Delta_\sigma$, $g_{l\sigma}$ are the coupling constants between light and atoms and $\Delta_{\sigma}$ is the atom-light detuning.
The operator $\h{F}_\sigma$ describes the coupling between the atomic density $\h{n}_\sigma(\b{r})=\h{\Psi}^\dagger_\sigma(\b{r}) \h{\Psi}_\sigma(\b{r})$ and the light field: since different spin states couple to different (circular) polarization one has 
\begin{equation}
\h{F}_\sigma = \int u^*_{1}(\b{r}) u_{0}(\b{r}) \h{n}_\sigma(\b{r}) \d \b{r}
\end{equation}
where $u_{l}(\b{r})$ is the mode function of the light mode $l$ ($l=0,1$). This property has been exploited for investigating spin-spin correlations and magnetic susceptibility of a Fermi gas in previous works where the effect of measurement backaction\cite{Roscilde2009,Eckert2008,Meineke2012,Sanner2012} or the atomic dynamics \cite{atoms} were neglected.  Expanding the matter field $\h{\Psi}^\dagger_\sigma(\b{r})$ ($\h{\Psi}_\sigma(\b{r})$) in terms of the lattice Wannier functions $w(\b{r})$ and the ladder operators of the lattice, we find that $\h{F}_\sigma=\h{D}_\sigma + \h{B}_\sigma$ where
\begin{equation}
\h{D}_\sigma =\sum_{i} J_{ii} \h{n}_{i\sigma} \qquad \h{B}_\sigma =\sum_{\langle i,j \rangle} J_{ij} \h{f}^\dagger_{i\sigma} \h{f}_{j\sigma}
\end{equation}
and 
\begin{equation}
J_{ij}=\int w(\b{r}- \b{r}_i) u^*_{1}(\b{r}) u_{0}(\b{r}) w(\b{r}- \b{r}_j) \d \b{r}.
\end{equation}
The operators $\h{D}_{\sigma}$ and $\h{B}_{\sigma}$ identify the two main contributions to the light scattering: the first one is due to the atomic density at each lattice site while the second depends on the inter-site coherence terms between neighboring sites.  If the atoms are well-localized, the coherence term can be usually neglected and $\h{F}_\sigma\approx\h{D}_\sigma$. However, carefully choosing the mode functions of the light, it is possible to suppress the on-site contribution to the light scattering so that the coupling  described by $\h{B}_{\sigma}$ dominates \cite{Kozlowski2015,Mazzucchi2015}. In \cite{Caballero2015,Caballero2015b,Caballero2016,Caballero2016bond} bosonic systems were considered in which the cavity mediates long-range interactions that are dominant and triggers emergent quantum  phases of matter. Here measurement effects were not analysed. In this work, we ignore the effect of cavity backaction and focus solely on the effect of measurement backaction since we consider the regime $\kappa\gg\Delta_p$ where $\Delta_p$ is the cavity-probe detuning.

From the Hamiltonian \eqref{eq:coupling} and assuming that $\kappa\gg J$ we compute the Heisenberg equations for the light field in the stationary limit.  Neglecting the cavity dispersion shift,  the amplitude of the scattered light  depends on the atomic density at each lattice site, in analogy to the classical case. The light field operator can be expressed as $\h{a}_1= \sum_\sigma C_\sigma \h{D}_{\sigma}$ where $C_{\sigma}$ is the Rayleigh scattering coefficient in presence of a cavity~\cite{Kozlowski2015} and it is given by
\begin{equation}
C_\sigma=\frac{i U_\sigma a_{0\sigma}}{i \Delta_p - \kappa}.
\end{equation}
The polarization of the probe beam defines which linear combination of the  $\uparrow$ and $\downarrow$ are addressed by the measurement scheme. For example, if the probe laser is circularly polarized ($L$ or $R$), the measurement process is sensitive only to one of the two spin species ($\h{a}_1=C_L \h{D}_\uparrow$ or $\h{a}_1=C_R \h{D}_\downarrow$). Furthermore, considering the case of linearly polarized probe, the photons escaping the optical cavity carry information about the atomic density $\h{\rho}_i=\h{n}_{i\up}+\h{n}_{i\down}$ and the magnetization $\h{m}_i=\h{n}_{i\up}-\h{n}_{i\down}$ so that the annihilation operators for the cavity field are proportional to
$\h{D}_x=\sum_{i=1}J_{ii} \h{\rho}_{i}$ ($x$-polarized light) and $\h{D}_y=\sum_{i=1}J_{ii} \h{m}_{i}$ ($y$-polarized light).

Since the photon annihilation operator $\h{a}_1$ depends on the matter observables, the photons escaping the optical cavity allow us to continuously monitor the quantum state of the system. In this Article, we focus on the conditional dynamics of the atoms in a single experimental realization, describing it with the quantum trajectories formalism~\cite{MeasurementControl}. The state of the system is a result of the deterministic evolution given by the non-Hermitian Hamiltonian $\h{H}_\mathrm{eff}=\h{H}_0- i \hbar \h{c}^\dagger \h{c} /2$  and the stochastic quantum jumps when the operator $\h{c}=\sqrt{2 \kappa} \h{a}_1$ is applied to the atomic state. We compute the conditional dynamics as follows: (i) a random number $r\in [0,1)$ is generated , (ii) the state of the system is propagated in time using the effective Hamiltonian $\h{H}_\mathrm{eff}$ until its norm reaches $r$ and then (iii) the quantum jump operator is applied to the atomic state which is subsequently normalized. This three-steps process is then repeated starting from (i). Note that the non-Hermitian term in $\h{H}_\mathrm{eff}$ is characterized by the energy scale $\gamma=\kappa |C|^2$ which competes with the usual tunneling amplitude $J$ and on-site interaction~$U$, leading to new many-body dynamics not described by the Hubbard Hamiltonian~\cite{Mazzucchi2015} and novel effects beyond the quantum Zeno limit \cite{Kozlowski2016nh} . 

\subsection*{Spatially-structured global coupling}

The coefficients $J_{ij}$ determines the spatial profile of the measurement operator, making our setup extremely flexible. Depending on the mode functions of the light modes and  on the angles between the optical lattice, the cavity and the probe, it is possible to address different observables. Focusing on a one-dimensional atomic chain along the $z$ axis and considering  traveling waves as mode functions (so that $u_l(\b{r})=e^{i \b{k} \cdot \b{r}}$) the coefficients $J_{jj}$ describe the usual diffraction from a periodic grating and are given by $J_{jj}=e^{i( \b{k}_0-\b{k}_1)\cdot \b{r}_j}=e^{i\delta j}$, where $\delta=( k_{0,z}-k_{1,z})d$ and the subscript $z$ represents the projection along the $z$ axis so that $k_{l,z}=|\b{k}_l| \cos \theta_l$, $l=0,1$. Simply adjusting the angles $\theta_0$ and $\theta_1$ allows us to probe and affect different linear combinations of the atomic density. If the value of $J_{jj}$ is the same on a subset of sites of the optical lattice, atoms in this region scatter light with the same phase and are therefore indistinguishable by the measurement~\cite{Elliott2015}. As a consequence, the detection process divides the optical lattice in different spatial modes composed by non-contiguous lattice sites that show long-range entanglement and correlations. Since the measurement process is only sensitive to global observables, the dynamics of the modes preserves quantum superpositions and can entangle distant lattice sites by enhancing specific dynamical processes~\cite{Mazzucchi2015}, relaxing the requirements for single-site or other spatial resolution necessary to obtain various important effects~\cite{Ashida2015, Wade2015}. With reference to the previous example, if $\delta=2 \pi s /R$ ($s,R \in \mathbb{Z}^+$) the lattice is partitioned in $R$ spatial modes since atoms separated by $R$ lattice sites scatter light with the same phase and amplitude, making them indistinguishable to the measurement. Therefore, the scattered light operator reduces from being a sum of numerous microscopic contributions from individual sites to the sum of smaller number of macroscopically occupied regions ($\h{a}_1 \propto \sum_{m=1}^R e^{i 2 \pi s m/R} \h{N}_{m \sigma} $).

In this work, we focus mainly on two different measurement schemes which partition the lattice in two spatial modes. The first one addresses the difference in occupation between odd and even lattice sites ($\h{D}_{\sigma}=\hat{N}_\mathrm{\sigma,odd}-\hat{N}_\mathrm{\sigma,even}$) and  is implemented using traveling or standing waves and detecting the scattered photons in the diffraction minimum orthogonal to the direction of the probe beam. This corresponds to the case where the cavity is placed along the lattice direction (i. e. $\theta_1=0$) and the probe beam is perpendicular to them (i. e. $\theta_0=\pi/2$) so that $J_{ii}=(-1)^i$.   The second scheme we consider probes the number of atoms at the odd sites ($\h{D}_{\sigma}=\hat{N}_\mathrm{\sigma,odd}$). It is realized with standing waves ($u_i(\b{r}) = \cos(\b{k}_i \cdot \b{r })$), crossed at such angles to the lattice that $\b{k}_0\cdot \b{r }$ is equal to $\b{k}_1\cdot \b{r }$ and shifted such that the even sites are placed in the maxima of the light interference pattern, while the even sites are positioned at the interference zeros and do not scatter light, so that $J_{ii}=1$ for $i$ odd while $J_{ii}=0$ for $i$ even. Note that both the measurement scheme we consider address the system globally and do not require single-site resolution as the coefficients $J_{ii}$ are determined by the projections of the probe/cavity wave vectors on the direction of the optical lattice and not by their~wavelengths. By changing the angles $\theta_0$ and $\theta_1$ it is possible to implement measurement operators with different spatial profiles as the periodicity of the coefficients $J_{ij}$ can be easily made larger (not smaller) than the lattice period.

\begin{figure}[t!]
\centering
\includegraphics[width=0.6\textwidth]{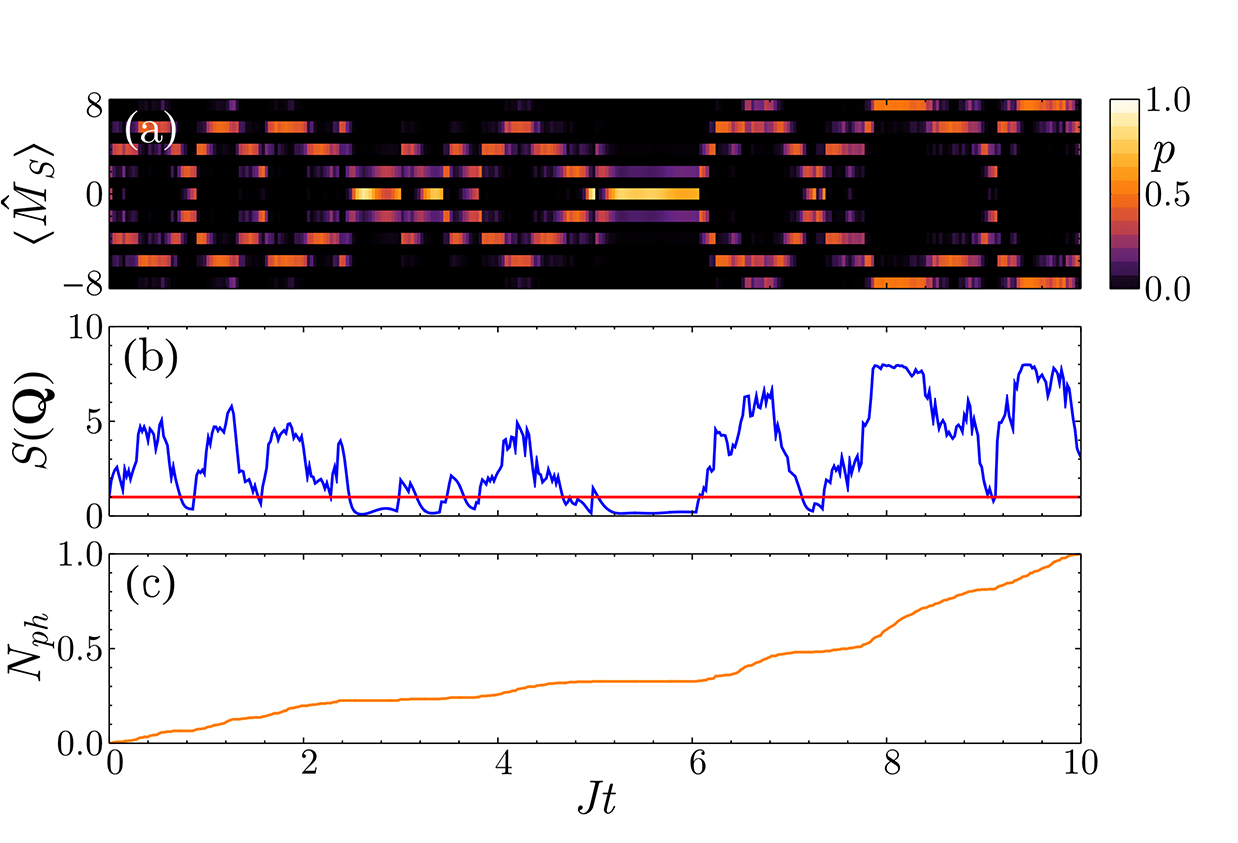}
\caption{\textbf{Measurement-induced AFM order in a single experimental run.} (a) The probability distribution of the staggered magnetization $\m{\h{M}_s}$ presents two strong peaks as the state of the system is in a quantum superposition analogous to a Schr\"odinger cat state. (b) Comparison between the magnetic structure factor $S(\b{Q})$ for the ground state (red) and the conditional dynamics (blue), confirming the presence of AFM order. (c) Number of detected photons as a function of time (normalized to one). The derivative of this curve (photocount rate) is proportional to $S(\b{Q})$. ($\gamma/J=1$, $U/J=0$, $N_\up=N_{\down}=4$, $L=8$)} \label{min}
\end{figure}

\subsection*{Measurement-induced antiferromagnetic ordering}
We first focus on non-interacting fermions with two spin components at half filling ($N_{\up}=N_{\down}=L/2$) and we detect the light scattered in the diffraction minimum. In this case, the ground state of the system is the Fermi Sea $\ket{FS}$ where only single particle states with $k<k_F$ are occupied ($k_F$ being the Fermi wavevector) and the density and magnetization are uniform across the lattice. We use this state as a reference point, assuming that the atomic system is initialized in its ground state before the measurement take place. 
The monitoring process perturbs this state and induces AFM correlations that can drive the atomic state to a superposition of the Neel states $\ket{\up\down\up\down ....}$ and $\ket{\down\up\down\up ....}$. In contrast to previous works \cite{ritsch2013,Kramer2014,Morigi2015,Piazza2015, Caballero2015,Caballero2015b,Keeling2014superradiance,Piazza2014superradiance, Chen2014superradiance},this state emerges as a consequence of the competition between measurement backaction and atomic tunneling in a single quantum trajectory and does not rely on the cavity potential. Probing the atomic system with linearly polarized light along the $y$ axis, the annihilation operator describing the photons escaping the cavity is  $\h{a}_1=C\sum_{i=1} (-1)^{i} \h{m}_{i}=C(\h{M}_{\mathrm{even}}-\h{M}_{\mathrm{odd}})\equiv C \hat{M}_s$ so the measurement is directly addressing the staggered magnetization of the atomic system.  Moreover, the photon number operator $\h{a}^{\dagger}_{1} \h{a}_1$ does not depend on the sign of $\h{M}_s$ and therefore does not distinguish between states with opposite magnetization profile. Consequently, the conditional dynamics preserves $\m{\h{M}_s}$ and the local magnetization remains the same as the ground state ($\m{m_{i}}=0$ for all $i$). The quantum jumps tend to suppress states that do not present AFM correlations since the application of $\h{c}$ on the atomic state completely vanishes its components with $\m{\h{M}_s^2}=0$. Therefore, the detection process modifies the probability distribution of $\m{\hat{M}_s}$, making it bimodal with two symmetric peaks around $\m{\hat{M}_s}=0$ that reflects the degeneracy of $\h{a}^{\dagger}_{1} \h{a}_1$. The evolution of such peaks depends on the ratio $\gamma / J$ which determines whether the dynamics is dominated  by the quantum jumps or by the usual tunneling processes. Specifically, in the strong measurement regime ($\gamma \gg J$) the detection process freezes $\m{\h{M}_s^2}$ to a specific value stochastically determined by a particular series of quantum jumps. However, if $\gamma \ll J$ the measurement cannot inhibit the dynamics and the tunneling of atoms across the optical lattice leads to an oscillatory behavior, that, in the case of bosons, can be described analytically \cite{Mazzucchi2016b}.

The presence of AFM correlations in the quantum state resulting from the conditional dynamics is revealed by computing the magnetic structure factor 
\begin{equation}
S(\b{q})=\frac{1}{L}\sum_{i,j} e^{i \b{q} \cdot (\b{r}_{i}-\b{r}_{j})} \left( \m{\h{m}_{i} \h{m}_{j}} - \m{\h{m}_{i}} \m{\h{m}_{j}}\right).
\end{equation}
Monitoring this quantity for each quantum trajectory, we find that the measurement induces a strong peak at $\b{q}=\b{Q}=\pi/d$ and creates an AFM state (Fig.~\ref{min}). Importantly, the value of $S(\b{Q})$ is directly accessible to the experiments since the probability for a photon to escape the optical cavity in a (small) time interval $\d t$ is proportional to $ \m{\h{c}^\dagger \h{c}} \d t$. This allows to observe the formation of AFM ordering in real-time by simply computing the photocount rate.  Note that the emergence of these intriguing quantum states is stochastic  and varies depending on the specific quantum trajectory, i. e. a single experimental set of photodetections. However, thanks to the photons escaping from the cavity,  it is possible to precisely determined when the AFM correlations are established and any subsequent  dynamics can be frozen by increasing the depth of the optical lattice. The resulting state can then be used for more advanced studies with applications to quantum simulations and quantum information. In contrast to condensed matter systems or usual cold atoms experiments where a strong repulsion between the atoms is necessary for establishing AFM order, our measurement scheme allows to obtain these states even for non-interacting fermions. Furthermore, the spatial period of the correlations imprinted by the measurement can be tuned changing the scattering angle and may lead to the realization of states with more complex spatial modulations of the magnetization. Therefore, global light scattering will help to simulate effects of long-range interactions in Fermi systems which are inaccessible in contemporary setups based on optical lattices with classical light.

\begin{figure}[t!]
\centering
\includegraphics[width=0.6\textwidth]{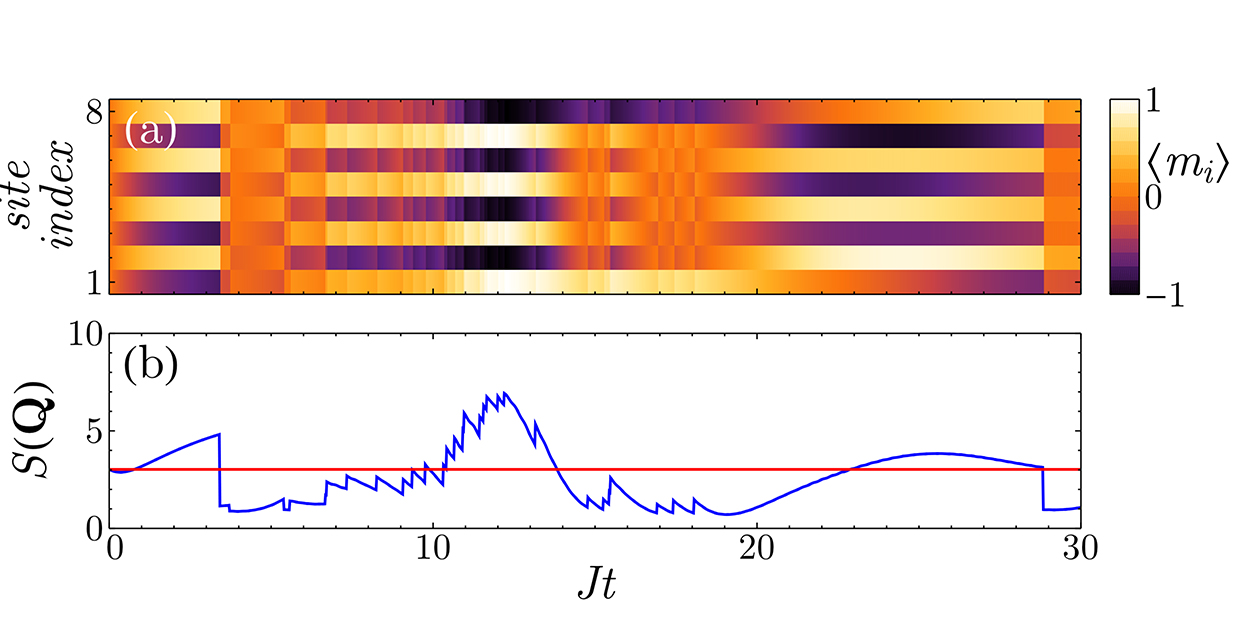}
\caption{\textbf{Measurement-induced AFM order in a single experimental run for repulsive fermions.} (a) The measurement process creates a modulation in the local magnetization, which is not present in the ground state. (b)Comparison between magnetic structure factor $S(\b{Q})$ for the ground state (red) and the conditional dynamics (blue), confirming the enhancement of AFM correlations. ($\gamma/J=0.1$, $U/J=20$, $N_\up=N_{\down}=4$, $L=8$)} \label{minU}
\end{figure}

Considering the case of interacting fermions, we show that addressing only one of the two spin species affects the global density distribution and induces AFM correlations.  Specifically, we consider the measurement operator $\h{D}_{L}=\h{N}_{\mathrm{odd}\up}$, which probes only one spin species using circularly polarized light, and we focus on the strongly repulsive limit $U=20J$. The detection process is sensitive only to the spin-$\up$ density and therefore induces a periodic modulation of the spatial distribution of this species, favoring either odd or even lattice sites. However, since the two different spins are coupled by the repulsive interaction, spin-$\down$ atoms are also affected by the measurement and tend to occupy the lattice sites that are not occupied by the spin-$\up$ atoms, enhancing the AFM correlations of the ground state. Importantly, in this case the  AFM character of the atomic state is visible in the local magnetization~(Fig.~\ref{minU}) since the photon number operator can distinguish between states with opposite values of $\m{\hat{M}_s}$. The presence of a single peak in the probability distribution of $\m{\hat{M}_s}$ makes this scheme more robust to decoherence due photon losses than detecting the scattered light in the diffraction minimum.

\subsection*{Measurement-induced density ordering}

\begin{figure}[t!]
\centering
\includegraphics[width=0.6\textwidth]{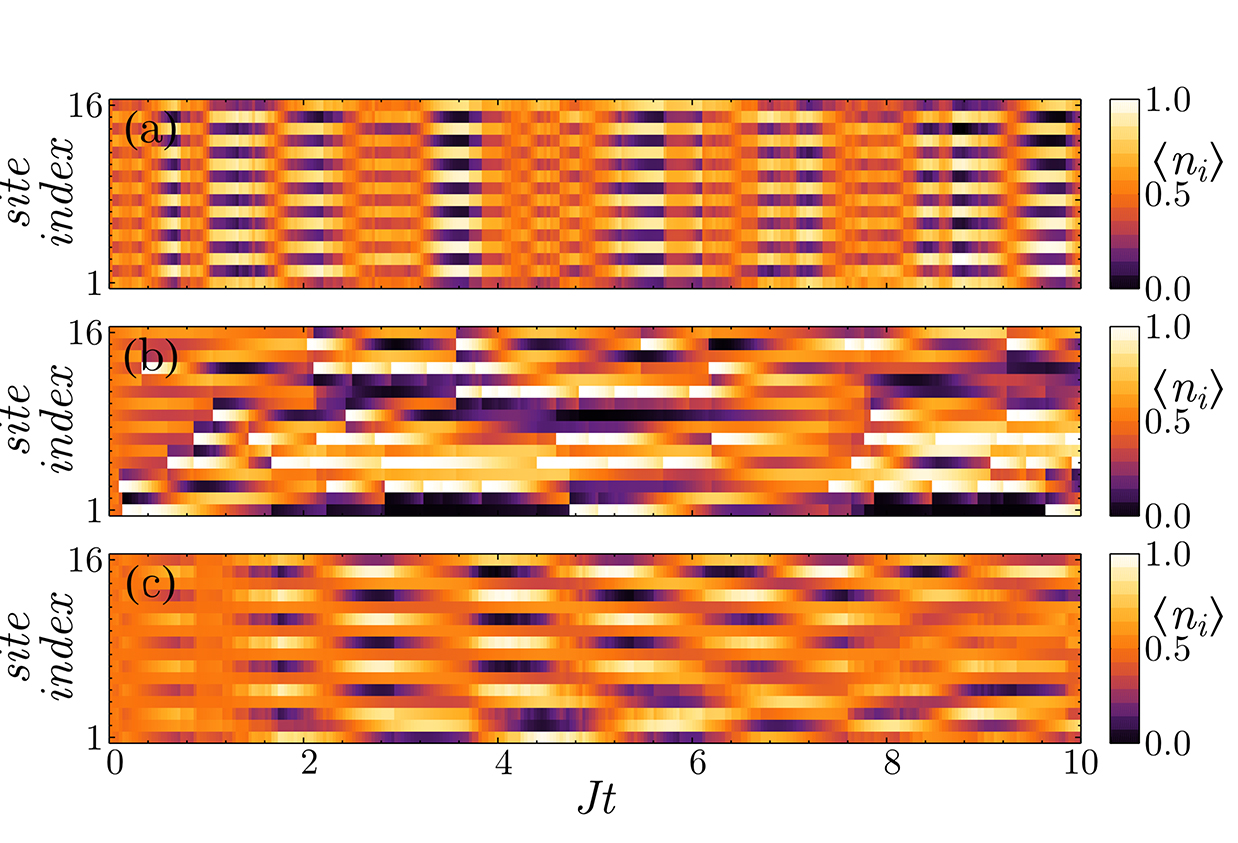}
\caption{\textbf{Conditional evolution of the local density in a single experimental run probing the optical lattice with (a,c) global and (b) local addressing illuminating the odd lattice sites.} (a)~The atomic population collectively oscillates between the two modes defined by the measurement establishing a spatial periodic modulation with period $2d$. (b)~The measurement process suppresses the local fluctuations independently for each lattice site.(c)~Changing the detection angle it is possible to tune the spatial period of the oscillations, establishing a density wave with period $3d$ ($\gamma/J=1$, $U/J=0$, $N=N_\up=8$, $L=16$)} \label{frozen}
\end{figure}

We now turn to non-interacting polarized fermions, i.~e. all the atoms are in the same spin state and the atomic Hamiltonian only describes tunneling processes between neighboring lattice sites.  The effective dynamics emerging by measuring $\hat{c}\propto\h{N}_\mathrm{odd}$  modulates the atomic density across the lattice and depends on the ratio $\gamma/J$. If the measurement is weak, the atoms periodically oscillate between odd and even sites so that $\m{\h{N}_{\mathrm{odd}}-\h{N}_{\mathrm{even}}}\neq 0$, i. e. a state where a density wave with the period  of the lattice is established. Such configuration is usually a consequence of finite range interactions~\cite{Hirsch1984} and it has been observed in solid state systems~\cite{PhysRevB.16.801} and molecules in layer geometries~\cite{Block2012}.  Here, in contrast, the atoms do not interact but the cavity coupling with all the lattice sites mediates an effective interaction between them. Note that global quantum nondemolition measurements have been already proposed for molecules in low dimensions \cite{MekhovLP2013}, which can link these fields even closer. Observing the photons leaving the cavity allows us to continuously monitor the state of the atoms, precisely determining when the density wave is established without need of external feedback~\cite{Ivanov2014,Pedersen2014,Wade2015}. 
If $\gamma \gg J$, the amplitude of the density wave remains constant on a timescale larger than $1/J$. Importantly, the fluctuation of the expectation value of the $\h{N}_{\mathrm{odd}}$ are strongly suppressed even if the on-site atomic density is not well-defined. This is a consequence of the global addressing of our measurement scheme: the jump operator does not distinguish between different configurations having the same $\h{N}_\mathrm{odd}$. This property is crucial for establishing correlations between distant lattice sites (Figure~\ref{frozen}). Local~\cite{Diehl2008} or fixed-range~\cite{LesanovskyPRL2012} addressing destroys coherence between different lattice sites and tends to project the atomic state to a single Fock state, failing to establish a density wave with a well-defined spatial period. Furthermore, the spatial period of the long-range correlations imprinted by the global measurement can be easily tuned changing the scattering angle. We illustrate this by considering the coefficients $J_{ii}=[1,\, 1/2,\, 0,\, 1/2,\, 1,\, 1/2,\, 0...]$ which can be obtained using standing waves crossing the lattice at an angle such that  $\b{k}_{\mathrm{1,0}}\cdot \b{r}=\pi/4$. With this scheme, the measurement partitions the lattice in $R=3$ non-overlapping spatial modes and leads to the emergence of density modulations with period~$3d$~(Figure~\ref{frozen}c).

\begin{figure}[t!]
\centering
\includegraphics[width=0.7\textwidth]{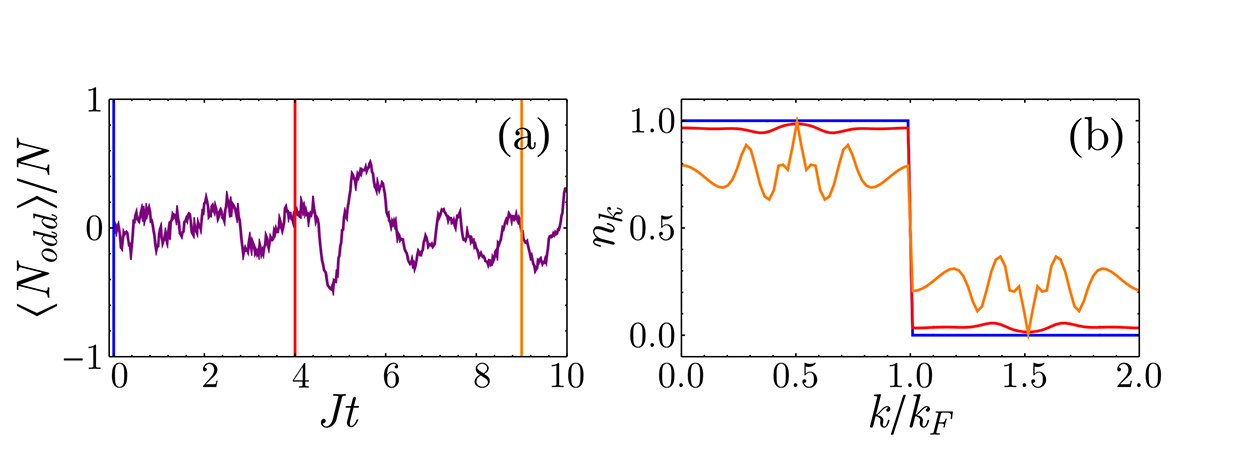}
\caption{\textbf{ Solution of the mean field equations for a single experimental run.} (a) The measurement imprints a density modulation on the atomic state as the number of atoms occupying the odd sites oscillates. (b) Occupation of the single particle momentum states for different times showing that the creation of particle-hole excitations is symmetric around $k_F$.  Different colors represent different times with reference to panel~(a) ($\gamma/J=0.05$, $U/J=0$, $N=50$, $L=100$)} \label{MF}
\end{figure}

The measurement backaction changes the structure of the atomic state taking the system away from its ground state. The specific form of the excitations on top of the Fermi Sea depends on the spatial structure of the jump operator $\h{c}$ and, changing the spatial profile of $J_{ii}$, allows us to select which momentum states are affected by the measurement process. Defining $A_\b{k}$ to be the Fourier transform of $A_i=\sqrt{2 \kappa} C J_{ii}$, one has 
\begin{equation}
\h{c}=\sum_{\b{k},\b{p}\in BZ} A_\b{p} \h{f}^\dagger_{\b{k}} \h{f}_{\b{k}+\b{p}},
\end{equation}
where BZ indicates the first Brillouin zone. If $A_\b{k}$ presents a narrow peak around $\b{k}=0$, i.~e. the measurement probes the number of atoms rather homogeneously in an extended region of the lattice ($J_{ii}\sim \mathrm{const}$), the detection process creates particles and holes only on the Fermi surface, which is the typical scenario in conventional condensed matter systems.  In contrast to this, the setup we propose allows us to probe states that are deep in the Fermi Sea: if $\h{c}\propto \h{N}_\mathrm{odd}$ one has $A_\b{k}\propto \delta(\b{k})+\delta(\b{k}+\b{Q})$ and the resulting jump operator is
\begin{equation}
\h{c}\propto \sum_{\b{k}\in BZ} \h{f}^\dagger_{\b{k}} \h{f}_{\b{k}}+\h{f}^\dagger_{\b{k}} \h{f}_{\b{k}+\b{Q}}.
\end{equation}
Applying this expression on the ground states leads to $\h{c} \ket{FS} \propto N \ket{FS}/2 + \ket{\Phi}$ where
\begin{equation}
\ket{\Phi}=\sum_{\b{k}:|\b{k}+\b{Q}|<k_F}\h{f}^{\dagger}_{\b{k}}\h{f}_{\b{k}+\b{Q}} \ket{FS}
\end{equation}
and $\braket{\Phi}{FS}=0$. Therefore, the detection process creates particle-hole excitations with momenta that are symmetric around the wavevector $\b{Q}/2$ and are not necessarily confined around $k_F$. This symmetry is reflected in the occupation of the single particle states (Figure~\ref{MF}b) and can be better understood defining $\h{\beta}_{\b{k}} = (\h{f}_{\b{k}} +\h{f}_{\b{k}+\b{Q}} )/\sqrt{2}$ and rewriting the jump operator as $\h{c}=\sum_{\b{k}\in RBZ} \h{\beta}_{\b{k}}^\dagger \h{\beta}_{\b{k}}$ where $RBZ$ is the reduced Brillouin zone. Therefore, the measurement tends to freeze the number of particle-hole excitations with wavevectors that are symmetric superposition around~$\b{Q}/2$.

The emergence of long-range entanglement makes the numerical solution of the conditional dynamics in a single quantum trajectory an extremely challenging problem which is difficult to solve efficiently even with methods such as Matrix Product States (MPS)~\cite{Schollwock}. While, in general, the system dynamics for small and large particle numbers can be indeed different \cite{mekhovLP2011}, here we confirm our findings on larger systems by formulating a mean field theory for the stochastic evolution of single particle  occupation number $n_{\b{k}}=\m{\h{f}^{\dagger}_{\b{k}} \h{f}_{\b{k}}}$ and the order parameter $\alpha_{\b{k}}=\m{\h{f}^{\dagger}_{\b{k}} \h{f}_{\b{k}+\b{Q}}}$. In general, the evolution of the observable $\h{O}$ conditioned to the outcome of the measurement follows a generalization of the Ehrenfest theorem.  Between two quantum jumps the dynamics is deterministic and follows 
\begin{equation}
\der{t} \m{{\h{O}}}= -i \m{[\h{H}_{0},\h{O}]} - \m{\{\h{c}^{\dagger}\h{c},\h{O}\}}+ 2 \m{\h{O}}\m{\h{c}^{\dagger}\h{c}}
\end{equation}
where $[\cdot,\cdot]$ ($\{\cdot,\cdot\}$) is the (anti)commutator. The photocurrent escaping the cavity follows a stochastic process where the photocounts are determined by the norm of the atomic wavefunction $\ket{\Psi}$ which is given by 
\begin{equation}
\der{t} {\braket{\Psi}{\Psi}}= -2 \m{\h{c}^{\dagger}\h{c}}
\end{equation}
When a photon is detected, the quantum jump operator is applied to the atomic state and the value of $\h{O}$ changes~as 
\begin{equation}
 \m{\h{O}} \rightarrow \frac{\m{\h{c}^{\dagger} \h{O} \h{c}}}{\m{\h{c}^{\dagger}\h{c}}}.
\end{equation}
We apply a mean field treatment to these equations decoupling the terms with more than two operators as a function of $n_{\b{k}}$ and $\alpha_{\b{k}}$ and we solved them numerically (Figure~\ref{MF}), confirming the emergence of density modulations even in large systems.

\subsection*{Inefficient detection}
The effects described in this work rely on the efficient detection ($\eta=1$) of the photons escaping the optical cavity. In this section, we show that the measurement induces AFM correlations and density ordering that can be observed even if $\eta<1$. In order to be able to determine if the magnetization (or density) of the atomic system presents a periodic spatial modulation, one needs to be able to estimate the photocurrent leaking from the optical cavity.  Focusing on the case illustrated in Fig.\ref{min}(c), the oscillations of the probability  distribution of $\m{\h{M}_s}$ are directly imprinted on the number of detected photons $N_{ph}$ which shows a ``staircase'' behavior. If this structure is resolved, the large-scale oscillations in the measured quantity survive and can be detected even if $\eta<1$. 

The effect of detection efficiency can be taken into account in our model by solving a stochastic master equation (SME) instead of performing a simple quantum trajectory calculation using a stochastic Schr\"odinger equation. Because of the undetected photons, the state of the system cannot be described by a pure state and it is necessary to introduce a density matrix representation. Specifically, given the jump operator $\hat{c}$ and the efficiency $\eta$, the density matrix $\h{\rho}$ of a system subjected to continuous inefficient monitoring evolves as \cite{MeasurementControl}
\begin{align}\label{sme}
\d \h{\rho} (t)= \left\{   \d N \mathcal{G} [ \sqrt{\eta}\hat{c} ] - \d t \mathcal{H} [i \hat{H}_0 + \frac{\eta}{2} \hat{c}^\dagger \hat{c} ]  + \d t (1 - \eta)\mathcal{D}[\hat{c}] \right\}\h{\rho}(t)
\end{align}
where $\d N$ is a stochastic It\^o increment such that $E[\d N]=\eta \Tr[\hat{c} \h{\rho} \hat{c}^\dagger]\d t$ and $\mathcal{G}, \mathcal{H}$ and  $\mathcal{D}$ are the superoperators
\begin{align}
\mathcal{G} [\hat{A} ]\h{\rho}=\frac{\hat{A}\h{\rho}\hat{A}^\dagger }{\Tr \left[\hat{A}\h{\rho}\hat{A}^\dagger \right]}-\h{\rho}\\
\mathcal{H} [\hat{A} ]\h{\rho}=\hat{A} \h{\rho} + \h{\rho} \hat{A}^\dagger - \Tr\left[ \hat{A} \h{\rho} + \h{\rho} \hat{A}^\dagger \right]\\
\mathcal{D} [\hat{A} ]\h{\rho}=\hat{A} \h{\rho} \hat{A}^\dagger - \frac{1}{2}\left(  \hat{A}^\dagger  \hat{A} \h{\rho} + \h{\rho}  \hat{A}^\dagger  \hat{A}\right).
\end{align}
Equation (\ref{sme}) reproduces the usual master equation if $\eta=0$ while it can be interpreted as stochastic Schr\"odinger equation (for pure states) if $\eta=1$. Because of the stochastic nature of the detection process, the function $N_{ph}(t)$ which describes the number of photon detected up to the time $t$, is given by
\begin{align}
N_{ph}(t)=\sum_{i=0}^{N_{ph}} \theta(t-t_i)
\end{align}
where $\theta(t)$ is the Heaviside step function and $t_i$ is the time at which the $i-$th (detected) jump takes place. However, if the average time between two quantum jumps is much smaller than the typical timescale of the atomic dynamics (i. e. $1/2\eta \langle \hat{c}^\dagger \hat{c}\rangle\ll 1/J$), the function $N_{ph}(t)$ approximately follows the differential equation
\begin{align}\label{der}
\frac{\d N_{ph}}{\d t}=\eta \langle \hat{c}^\dagger \hat{c}\rangle(t)
\end{align}
where the expectation value $\langle \dots \rangle$ is computed on a single realization of the stochastic process described by the It\^o increment $\d N$. If the jump operators is given by $\h{M}_s$ or $\h{N}_\mathrm{odd}$, the value of these quantities and their evolution is directly imprinted on the profile of $N_{ph}(t)$. Focusing on a single quantum trajectory for the density matrix $\h{\rho}$ and defining $N_e(t)$ as the number of photons that escape the cavity up to the time $t$ (and that are not necessarily detected) one has that the statistical properties of a single realization of $N_{ph}(t)$ follow the so-called Bernoulli process (each photon can be detected with probability $\eta$) and one has
\begin{align}
E[N_{ph}(t)]=\eta N_e(t),\\  Var[N_{ph}(t)]=\eta (1-\eta) N_e(t),
\end{align}
where $E$ is the mean value and $Var$ is the variance. From these expressions, we can estimate the signal-to-noise ratio (SNR) for the conditional dynamics of a quantum system that is continuously monitored using a detector with efficiency $\eta$:
\begin{align}\label{SNR}
SNR[N_{ph}]\sim\sqrt{\frac{\eta}{1-\eta}} \sqrt{ N_e(t)}
\end{align}
We can now infer what is the minimum efficiency required for determining the value of $\langle \hat{c}^\dagger \hat{c}\rangle(t)$ from the number of detected photons noting that the SNR of $\langle \hat{c}^\dagger \hat{c}\rangle(t)$ is analogous to equation (\ref{SNR}). Crucially, the measurement scheme we consider in this manuscript partitions the optical lattices in spatial modes with macroscopic occupation so that the value of  photon scattering rate $\langle \hat{c}^\dagger \hat{c}\rangle$ scales as the square of the particle number that are present in the optical lattice ($\sim \gamma N^2$). Thus, the number of photons $N_e$ that escape the cavity during the timescale defined by the tunneling ($2\pi/J$) scales as  $\gamma N^2 / J$. Therefore, based on Eq. (\ref{SNR}), the minimum efficiency required for having a good SNR is
\begin{align}
\eta\gtrsim \frac{J}{\gamma  N^2},
\end{align}
making this setup robust to detection inefficiency. Because of the relatively large amount of photons scattered \emph{collectively} by the atomic system \emph{as a whole} and thanks to the fact that the creation of quantum states with AFM correlations or density modulations does not rely on the detection of a \emph{single} photon, the effects that we describe in our manuscript are visible even if the detector has imperfect efficiency.

The physical insight in this estimation can be formulated as follows. Crucially, the growth of oscillation amplitude in the magnetization or density (i.e. appearance of AFM order or density modulation) is related to the corresponding oscillations in the measurement backaction. More precisely, the rate of photodetections (and thus the measurement backaction) should show well-pronounced oscillations with the tunneling frequency $J$, even if the detection efficiency is not perfect. Thus, this is a physical requirement for the measurement backaction to be still able to create the quantum state of interest. Instead of the oscillating photodetection rate, it is more convenient to deal with the number of detected photons, which instead of oscillations shows a growing staircase behavior with a characteristic time $2\pi/J$ (shown in Fig. 2). Thus, as long as the characteristic steps in the ``staircase'' are resolved, the measurement-based preparation is expected to be efficient. 

\section*{Discussion}

We have shown that measurement backaction on ultracold Fermi gases can be used for realizing intriguing quantum states characterized by a periodic spatial modulation of the density and the magnetization. We have demonstrated that this spatial structure is a consequence of the global nature of the coupling between atoms and light. The competition between measurement backaction and usual dynamics determined by the tunneling enables the study of quantum magnetism without requiring extreme cooling or strong interactions between the atoms.  Our method enables the possibility to engineer otherwise low entropy states (i.e. with small number of defects). Importantly, the formation of magnetic states can be observed in real-time by measuring the photocurrent leaving the optical cavity, without the need of destructive techniques such as time-of-flight imaging. Additionally, it might be possible to engineer feedback control to stabilise a desired correlated quantum many-body state. This opens new possibility for studying the dynamics of strongly correlated materials and the effect of magnetic ordering on superconducting states \cite{Mitrano} as the methods we described could be used for imprinting magnetic correlations on states with superconducting properties.  Moreover, the setup we presented is extremely flexible and, by tuning the spatial profile of the measurement operator, allows to realize different macroscopic quantum superpositions with applications ranging from quantum information to quantum technologies. It may be promising to extend recent phase-space methods~\cite{Ruostekoski2013} to fermionic systems~\cite{Rosales2015,Dalton2013} and to investigate the effect of measurement on atomic systems with higher spin.  Although light scattering into the cavity constitute an additional source of heating and decoherence for the atoms, the effects described in this work can be observed in present experimental setups~\cite{Hemmerich2015,Esslinger2015} since the life time of the important states is large enough and its dynamics can be probed~\cite{Landig2015}.  Moreover, ultracold bosons have been successfully trapped in optical cavities without a lattice~\cite{Landig2015,HemmerichScience2012,Bux2013} and light scattered from ultracold atoms loaded in an optical lattice without a cavity has been detected~\cite{KetterlePRL2011,Weitenberg2011}. Finally, Bragg spectroscopy of a cloud of ultracold bosons coupled to vacuum field of a cavity~\cite{Landig2015} and measurement-induced creation of Schr\"odinger cat states involving few thousand atoms~\cite{Vuletic2015} have been recently reported.

\bibliography{references}

\section*{Acknowledgements}

The work was supported by the EPSRC (EP/I004394/1).

 



\end{document}